\documentclass[aip,rsi,amsmath,amssymb,reprint,superscriptaddress]{revtex4-1}
\usepackage{graphicx}
\usepackage{dcolumn}
\usepackage{bm}
\usepackage{color}
\begin{document}
\title{Magnetic field sensitivity and decoherence spectroscopy of an ensemble of narrow-linewidth nitrogen-vacancy centers close to a diamond surface}
\author{Kento Sasaki}
\affiliation{School of Fundamental Science and Technology, Keio University, 3-14-1 Hiyoshi, Kohoku-ku, Yokohama 223-8522, Japan}
\author{Ed E. Kleinsasser}
\affiliation{\mbox{Department of Electrical Engineering, University of Washington, Seattle, Washington 98195-2500, USA}}
\author{Zhouyang Zhu}
\affiliation{\mbox{HKU-Shenzhen Institute of Research and Innovation (HKU-SIRI), Shenzhen 518000, China}}
\affiliation{\mbox{Department of Mechanical Engineering, The University of Hong Kong, Pokfulam, Hong Kong, China}}
\author{Wen-Di Li}
\affiliation{\mbox{HKU-Shenzhen Institute of Research and Innovation (HKU-SIRI), Shenzhen 518000, China}}
\affiliation{\mbox{Department of Mechanical Engineering, The University of Hong Kong, Pokfulam, Hong Kong, China}}
\author{Hideyuki Watanabe}
\affiliation{Correlated Electronics Group, Electronics and Photonics Research Institute, National Institute of Advanced Industrial Science and Technology (AIST),
Tsukuba Central 5, 1-1-1 Higashi, Tsukuba, Ibaraki 305-8565, Japan}
\author{Kai-Mei C. Fu}
\affiliation{\mbox{Department of Electrical Engineering, University of Washington, Seattle, Washington 98195-2500, USA}}
\affiliation{\mbox{Department of Physics, University of Washington, Seattle, Washington 98195-1560, USA}}
\author{Kohei M. Itoh}
\email{kitoh@appi.keio.ac.jp}
\affiliation{School of Fundamental Science and Technology, Keio University, 3-14-1 Hiyoshi, Kohoku-ku, Yokohama 223-8522, Japan}
\affiliation{\mbox{Spintronics Research Center, Keio University, 3-14-1 Hiyoshi, Kohoku-ku, Yokohama 223-8522, Japan}}
\author{Eisuke Abe}
\email{e-abe@keio.jp}
\affiliation{\mbox{Spintronics Research Center, Keio University, 3-14-1 Hiyoshi, Kohoku-ku, Yokohama 223-8522, Japan}}
\date{\today}

\begin{abstract}
We perform pulsed optically detected electron spin resonance to measure the DC magnetic field sensitivity and
electronic spin coherence time $T_2$ of an ensemble of near-surface, high-density nitrogen-vacancy (NV) centers engineered to have a narrow magnetic resonance linewidth.
Combining pulsed spectroscopy with dynamic nuclear polarization, we obtain the photon-shot-noise-limited DC magnetic sensitivity of 35~nT~Hz$^{-0.5}$.
We find that $T_2$ is controlled by instantaneous diffusion, enabling decoherence spectroscopy on residual nitrogen impurity spins in the diamond lattice and
a quantitative determination of their density.
The demonstrated high DC magnetic sensitivity and decoherence spectroscopy are expected to broaden the application range for two-dimensional magnetic imaging.
\end{abstract}
\maketitle

Submicron scale, two-dimensional (2D) magnetic imaging has potential applications in biological and physical sciences.~\cite{GM02,HO05}
The realization utilizing an ensemble of nitrogen-vacancy (NV) centers in diamond is particularly attractive
due to its high magnetic sensitivity at ambient conditions.~\cite{SCLD14,MWF+10,SDN+10,PLS+11,LAG+12,GLP+15,CTA+15,STM+16}
When NV-based sensing is carried out with continuous wave (CW) optically detected magnetic resonance (ODMR),
the photon-shot-noise-limited DC magnetic field sensitivity $\eta_{\mathrm{sn}}$ is estimated as
\begin{equation}{
\eta_{\mathrm{sn}}^{(\mathrm{cw})} = \frac{h}{g \mu_B} \frac{\delta \nu}{C \sqrt{I_0}},
\label{eta_cw}}
\end{equation}
where $h/g \mu_{\mathrm{B}}$ = 36~$\mu$T/MHz is the inverse of the gyromagnetic ratio of the NV electronic spins,
$I_0$ is the count rate of photons from the NV centers in a unit area (1~$\mu$m$^2$) under the off-resonance condition,
$\delta \nu$ is the ODMR linewidth,
and $C$ is the ODMR contrast (the ratio of the photon counts on and off resonance).~\cite{TCC+08,RTH+14}
Equation~(\ref{eta_cw}) suggests that simultaneously achieving a high NV density and a narrow linewidth is desired to improve $\eta_{\mathrm{sn}}^{(\mathrm{cw})}$.
In addition, the NV sensor must be located as close as possible to a magnetic specimen;
it is crucial to have an NV ensemble near the diamond surface.~\cite{WJGM13,ORW+13}

Recently, some of the present authors have reported the successful creation of a 100-nm-thick layer of NV ensembles at a diamond surface 
with the density of 10$^{17}$~cm$^{-3}$ and $\delta \nu$ of $\sim$200~kHz.~\cite{KSB+16}
This was achieved by a combination of chemical vapor deposition (CVD) growth of nitrogen-doped, nuclear-spin-free $^{12}$C (99.9\%) diamond film~\cite{IW14}
and subsequent helium ion implantation to introduce vacancies into the film.
The detailed procedure for the NV formation as well as the characterization of the NV ensemble by photoluminescence (PL) spectroscopy and CW ODMR are given in Ref.~\onlinecite{KSB+16}.

In this paper, we show that, by concurrently applying pulsed ODMR and dynamic nuclear polarization (DNP) techniques to the same diamond sample,
it is possible to realize $\eta_{\mathrm{sn}}$ of 35~nT~Hz$^{-0.5}$.
We also examine coherence properties of the NV ensemble to extract quantitative information on residual paramagnetic impurities in the sample.
This is an example of ``decoherence spectroscopy'', in which magnetic signals are detected via the change in spin coherence time $T_2$.~\cite{GBN+11,LSB+12,MSR12,GWD+14,LGH+15}
The method is applicable to identify magnetic signals external to the sample, providing another versatile tool for ensemble-based 2D magnetic field imaging.

\begin{figure}
\begin{center}
\includegraphics{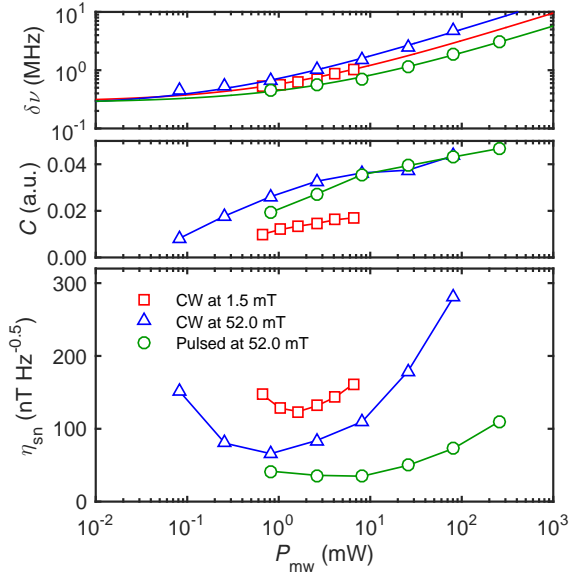}
\caption{
$\delta \nu$ (top), $C$ (middle), and $\eta_{\mathrm{sn}}$ (bottom) as functions of $P_{\mathrm{mw}}$ measured at the input port of the PCB board on which the sample is mounted.
The solid lines in the top panel are fits by $\delta \nu_0 + a P_{\mathrm{mw}}^{0.5}$.
The laser power $P_{\mathrm{L}}$ was optimized as 100~$\mu$W and 1.4~mW for CW and pulsed experiments, respectively.
\label{fig1}}
\end{center}
\end{figure}

We first recap the main result of Ref.~\onlinecite{KSB+16} by performing CW ODMR at the external magnetic field $B_0$ of 1.5~mT.
The squares ($\square$) in Fig.~\ref{fig1} are the measured $\delta \nu$ (top) and $C$ (middle) together with $\eta_{\mathrm{sn}}^{(\mathrm{cw})}$ estimated from Eq.~(\ref{eta_cw}) (bottom)
as functions of the microwave power $P_{\mathrm{mw}}$, demonstrating the minimum sensitivity of 124~nT~Hz$^{-0.5}$.
The measurement setup in the present work is a home-built confocal microscope combined with microwave circuitry, enabling CW and pulsed ODMR of single and ensemble NV centers.
Throughout this work, $B_0$ is applied parallel to one of four NV axes, and
the $m_S = 0 \leftrightarrow -1$ transition of the NV ensemble aligned to the field is examined, unless otherwise mentioned.
The relative improvement in $\eta_{\mathrm{sn}}^{(\mathrm{cw})}$ over the previous result (170~nT~Hz$^{-0.5}$ in Ref.~\onlinecite{KSB+16})
is attributed to the differences in collection efficiency and measurement location within the sample.

As is evident from Eq.~(\ref{eta_cw}), the magnetic sensitivity can be improved by optimizing $\delta \nu$, $C$ and $I_0$. 
Indeed, the low-field data in Fig.~\ref{fig1} show an interplay between $\delta \nu$ and $C$;
the narrowing of $\delta \nu$ as decreasing $P_{\mathrm{mw}}$ is countered by the reduction of $C$,
and $\eta_{\mathrm{sn}}^{(\mathrm{cw})}$ takes its minimum at an intermediate value of $P_{\mathrm{mw}}$ = 1.64~mW.
We observe that $\delta \nu$ behaves as $\delta \nu_0 + a P_{\mathrm{mw}}^{0.5}$ with $\delta \nu_0 \approx$ 250~kHz [solid line in the top panel of Fig.~\ref{fig1}].
On the other hand, $C$ and $I_0$ can be further improved, respectively, by employing DNP of $^{14}$N nuclei ($I$ = 1) associated with the NV centers and pulsed ODMR.

\begin{figure}
\begin{center}
\includegraphics{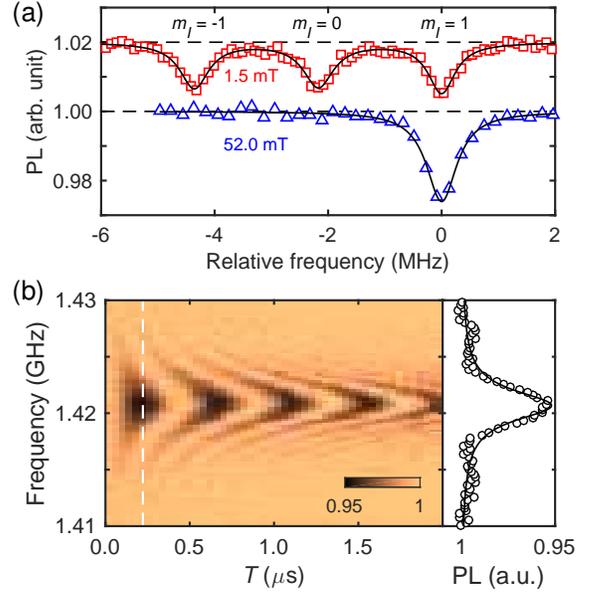}
\caption{
(a) CW ODMR spectra at $B_0$ = 1.5~mT, $P_{\mathrm{mw}}$ = 1.64~mW (shifted upward by 0.02 for clarity) and at $B_0$ = 52.0~mT, $P_{\mathrm{mw}}$ = 0.82~mW.
(b) An example of pulsed spectroscopy.
The cross section at $T_{\pi}$ = 222~ns is fitted by a Lorentzian.
\label{fig2}}
\end{center}
\end{figure}

It is well-established that the excited state of the NV center experiences a level anticrossing near 50~mT.
In this condition, optical pumping of the NV electronic spins polarizes the $^{14}$N nuclei into the $m_I$ = 1 state owing to electron-nuclear flip-flops.~\cite{JNB+09,FJKB13}
Figure~\ref{fig2}(a) plots CW ODMR spectra taken at 1.5~mT and 52.0~mT, demonstrating clear DNP in the latter.
We then repeat the measurements at 52.0~mT [$\bigtriangleup$ in Fig.~\ref{fig1}] to obtain the minimum sensitivity of 66~nT~Hz$^{-0.5}$ at $P_{\mathrm{mw}}$ = 0.82~mW.

In the DNP condition, by pumping the three nuclear spin states into a single state, we should ideally achieve a factor-of-three enhancement of $C$ and thus $\eta_{\mathrm{sn}}^{(\mathrm{cw})}$.
However, at a given $P_{\mathrm{mw}}$, $C$ under the DNP is typically only twice as deep as that at low fields.
Also, the values of $C$ giving the minimum sensitivities are 1.5\% at 1.5~mT and 2.6\% at 52.0~mT [Fig.~\ref{fig2}(a)], which are reflected in the obtained sensitivities (124~nT~Hz$^{-0.5}$ vs. 66~nT~Hz$^{-0.5}$).
$C$ is determined by a complicated interplay between $T_1$, $T_2^{\ast}$ and other optical transition probabilities between the NV electronic energy levels ({\it e.g.,} see Eq.~(A5) of Ref.~\onlinecite{DLR+11}).
We have examined several physical parameters of our NV ensemble to reproduce the observed $C$, but have not reached a satisfactory explanation.
We leave a detailed analysis on this as a future work.

In CW ODMR at a fixed $P_{\mathrm{mw}}$, increasing optical excitation power simultaneously increases $I_0$ and $\delta \nu$ while decreasing $C$.~\cite{DLR+11}
This leads to an optimal optical power well below the saturation intensity of the NV center. 
On the other hand, pulsed ODMR temporally separates the optical pumping from the spin manipulation.
A higher laser power can be used to significantly increase $I_0$ while keeping $C$ and $\delta \nu$ intact.~\cite{DLR+11}
An example of pulsed spectroscopy, $C$ as a function of the microwave burst time ($T$) and the microwave frequency, is shown in Fig.~\ref{fig2}(b).
We denote the experimental sequence as $\tau_{\mathrm{I}} - T - \tau_{\mathrm{R}}$,
where $\tau_{\mathrm{I, R}}$ are the durations of green laser excitation for spin initialization and readout.
By varying the microwave frequency around the $m_I$ = 1 resonance, a chevron pattern typical of pulsed spectroscopy is observed.
The cross section at the $\pi$ pulse condition ($T_{\pi}$ = 222~ns) is also shown in Fig.~\ref{fig2}(b), from which we deduce $\delta \nu$ and $C$ [$\bigcirc$ in Fig.~\ref{fig1}].
$\eta_{\mathrm{sn}}$ for pulsed ODMR is given by~\cite{DLR+11,KSB+16}
\begin{equation}{
\eta_{\mathrm{sn}}^{(\mathrm{pulsed})} = \frac{h}{g \mu_B} \frac{\delta \nu}{C} \sqrt{\frac{(\pi \delta \nu)^{-1} + \tau_{\mathrm{I}} + \tau_{\mathrm{R}} }{\tau_{\mathrm{R}} I_0}},
\label{eta_pulse}}
\end{equation}
and we obtain the minimum sensitivity of 35~nT~Hz$^{-0.5}$ for our system.

Through time-resolved fluorescence measurements (data not shown), we optimized $\tau_{\mathrm{I}}$ and $\tau_{\mathrm{R}}$ as 4.5~$\mu$s and 1.5~$\mu$s, respectively.
These spin initialization and readout times, several times longer than the case of a single NV center (typically about 1~$\mu$s and a few 100~ns, respectively),
currently limit the achievable $\eta_{\mathrm{sn}}^{(\mathrm{pulsed})}$.
The long $\tau_{\mathrm{I,R}}$ are attributed to the Gaussian profile of the laser spot.
Calculations suggest that 58~\% of the total fluorescence intensity arise from the region outside of the FWHM of the profile,
and the NV ensemble existing in this region is subject to substantially lower laser power,
resulting in insufficient initialization for shorter $\tau_{\mathrm{I}}$.
A long initialization time due to the laser spot profile has also been discussed in Ref.~\onlinecite{WNN+15}.

\begin{figure}
\begin{center}
\includegraphics{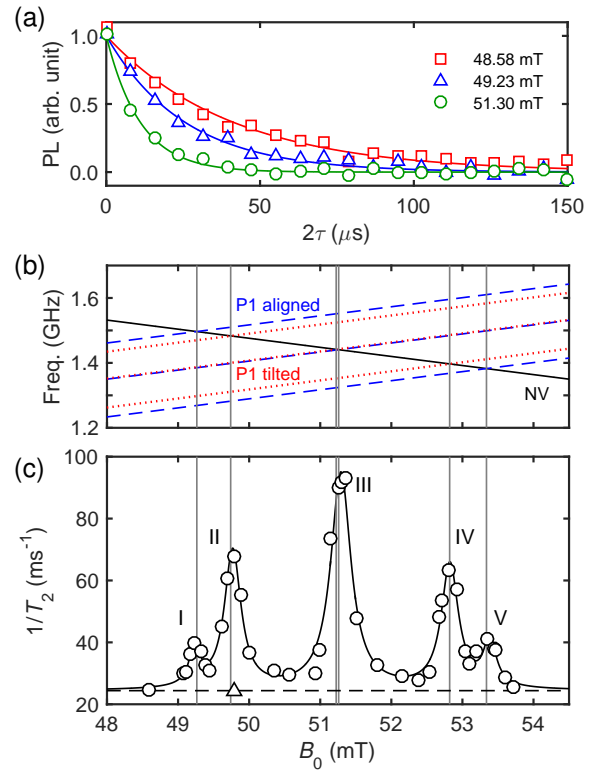}
\caption{
(a) Hahn echo decays of the NV ensemble at around $B_0$ = 50~mT.
The solid lines are fits by single-exponential decays.
(b) The transition frequencies of the NV centers (solid line, $m_S$ = 0 $\leftrightarrow$ $-$1 with $m_I$ = 1),
the P1 centers aligned with $B_0$ (dashed lines) and tilted from $B_0$ (dotted lines).
The unit and scale of the horizontal axis are the same as (c).
The vertical lines indicate $B_0$ at which the SSRs occur.
(c) $T_2^{-1}$ as a function of $B_0$.
\label{fig3}}
\end{center}
\end{figure}

Having demonstrated the potential of this sample for 2D DC magnetic sensing,
we next measure $T_2$ using a Hahn echo sequence ($\tau_{\mathrm{I}} - T_{\pi/2} - \tau - T_{\pi} - \tau - T_{\pi/2} - \tau_{\mathrm{R}}$) and carry out ensemble-based decoherence spectroscopy.
Figure~\ref{fig3}(a) shows three representative Hahn echo decay curves at around 50~mT, all described well by single-exponential decays $A \exp(-2\tau/T_2)$.
A detailed $B_0$-dependence of the decoherence rate shown in Fig.~\ref{fig3}(c) reveals a multi-peak structure.
As depicted in Fig.~\ref{fig3}(b), the peak positions [listed in Table.~\ref{tab1}] coincide with the simultaneous spin resonances (SSRs) of the NV and P1 centers:
$S$ = $\frac{1}{2}$ substitutional nitrogen impurities with $C_{3v}$ symmetry.~\cite{SSGL59,HGA06}
\begin{table}
\caption{\label{tab1}
Analysis of Fig.~\ref{fig3}(c).
The peak positions are determined by a quintuple-Lorentzian fit.
$(T_{\mathrm{id}}^{(i)})^{-1}$ specifies $(T_{\mathrm{id}})^{-1}$ at the $i$th peak after subtraction of $(T_2^{\mathrm{base}})^{-1}$.
$N_i$ is the impurity density.
}
\begin{ruledtabular}
\begin{tabular}{llcr}
Peak & $B_0$ (mT) & $(T_{\mathrm{id}}^{(i)})^{-1}$ (ms$^{-1}$) & $N_i$ (10$^{17}$ cm$^{-3}$) \\
\hline
I & 49.23 & 12.6 & 0.19\\
II & 49.77 & 44.8 & 0.68\\
III & 51.29 & 70.0 &1.05\\
IV & 52.82 & 40.2 & 0.61\\
V & 53.38 & 14.8 & 0.22
\end{tabular}
\end{ruledtabular}
\end{table}
In a dipolarly-coupled electron spin system, the refocusing pulse flips resonant spins within its bandwidth,
instantly changing local dipolar magnetic fields felt by the individual NV electronic spins: a process known as the instantaneous diffusion (ID).
At the SSR, the NV decoherence is accelerated by the increased number of flipped spins.
In the case of a homogeneous electron-spin distribution, the ID decay has a form of single-exponential $\exp(-2\tau/T_{\mathrm{id}})$ with $T_{\mathrm{id}}$ given by~\cite{KA62,SJ01}
\begin{equation}{
\frac{1}{T_{\mathrm{id}}} = DN \sin^2 \left( \frac{\beta}{2} \right), \,\, \mathrm{with} \,\, D = \frac{\pi \mu_0 g^2 \mu_B^2}{9 \sqrt{3} \, \hbar}.
\label{Tid}}
\end{equation}
Here, $\mu_0$ is the vacuum permeability, $N$ is the density of spins rotated, and $\beta$ is the flip angle of the refocusing pulse.
We are thus able to extract the densities of the P1 spins from the increased decoherence rate at the respective peaks.

To do so, we first define the {\it baseline} decoherence rate $(T_2^{\mathrm{base}})^{-1}$
as $T_2^{-1}$ = 24.4~ms$^{-1}$ = (41.1~$\mu$s)$^{-1}$ obtained for the $m_S$ = 0 $\leftrightarrow$ 1 transition at 49.8~mT driven at 4.2658~GHz [$\triangle$ in Fig.~\ref{fig3}(c)].
This spin state shares the same decoherence mechanism with the $m_S$ = 0 $\leftrightarrow$ $-$1 one, except for the ID due to the P1 spins.
The main contributor to $(T_2^{\mathrm{base}})^{-1}$ is the ID among the NV spins,
which, for $N_{\mathrm{NV} \parallel B_0}$ = 0.25 $\times$ 10$^{17}$~cm$^{-3}$ (1/4 of the total NV density~\cite{KSB+16}), is estimated to be 20.1~ms$^{-1}$.
The rest (4.3~ms$^{-1}$) should come from (i) the spectral diffusion caused by flip-flops among the P1 spins, and (ii) the $T_1$ relaxation.
The lattice $^{13}$C nuclei play a negligible role owing to the isotope enrichment of $^{12}$C in this sample.

We then calculate the P1-induced $T_{\mathrm{id}}^{-1}$ after subtracting $(T_2^{\mathrm{base}})^{-1}$ from the measured $T_2^{-1}$, and the corresponding P1 densities, which are listed in Table~\ref{tab1}.
In Eq.~(\ref{Tid}), $\beta = \pi/\sqrt{2}$ is used, due to the smaller rotation angle for the $S$ = $\frac{1}{2}$ P1 spins relative to the $S$ = 1 NV spins.~\cite{SJ01}
The ratio $(N_{\mathrm{II}} + N_{\mathrm{IV}})/(N_{\mathrm{I}} + N_{\mathrm{V}})$ = 3.1 is close to 3: the ratio of the numbers of P1 centers tilted from and aligned with $B_0$.
On the other hand, $(N_{\mathrm{I}} + N_{\mathrm{II}} + N_{\mathrm{IV}} + N_{\mathrm{V}})/2$ = 0.85 $\times$ 10$^{17}$~cm$^{-3}$ is less than $N_{\mathrm{III}}$ = 1.05 $\times$ 10$^{17}$~cm$^{-3}$.
The discrepancy may indicate the presence of an additional $S$ = $\frac{1}{2}$ impurity with the density of 2 $\times$ 10$^{16}$~cm$^{-3}$.
We note that, in samples with the P1 density of $\sim$10$^{17}$~cm$^{-3}$, the P1-induced spectral diffusion of the order of a few 100~$\mu$s has been observed,~\cite{PBBL+12,BPJB+13}
consistent with our assignment of a-few-ms$^{-1}$ decoherence rate to this mechanism.

\begin{figure}
\begin{center}
\includegraphics{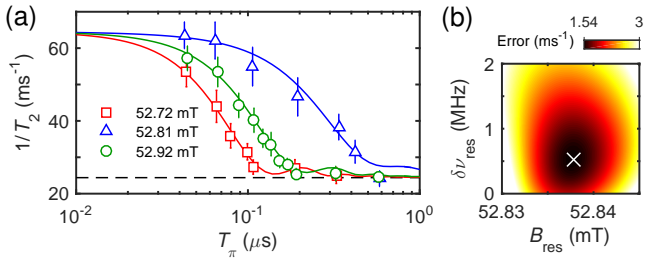}
\caption{
(a) $T_{\pi}$-dependence of $T_2^{-1}$ at Peak~IV.
(b) Fit error as a function of fitting parameters $B_{\mathrm{res}}$ and $\delta\nu_{\mathrm{res}}$.
The cross mark ($\times$) indicates the values giving the best fit.
\label{fig4}}
\end{center}
\end{figure}

Lastly, we examine the $T_{\pi}$-dependence of $T_2$, taking Peak~IV as an example.
$T_{\pi}$ of $\sim$45~ns used in Fig.~\ref{fig3}(c) is so broadband that the linewidth deduced from the multi-Lorentzian fit (16~MHz) does not necessarily reflect the true width.
Figure~\ref{fig4}(a) demonstrates that, as making $T_{\pi}$ longer, $T_2^{-1}$ gradually falls down to $(T_2^{\mathrm{base}})^{-1}$,
suggesting that less and less P1 spins are flipped.
Considering that an {\it effective} amount of spins rotated is determined from
an overlap between the frequency spectrum of the microwave pulse $\mathcal{P} = (f_{\mathrm{R}}/f_{\mathrm{R,g}})^2$ and the impurity spin spectrum $\mathcal{S}$,
we write $T_{\pi}$-dependent $T_2$ as
\begin{equation}{
\frac{1}{T_2} = \frac{1}{T_2^{\mathrm{base}}} + DN_{\mathrm{IV}} \int \mathcal{PS} \sin^2\left( \frac{\beta}{2} \right) df.
\label{Tcalc}}
\end{equation}
Here,
$f_{\mathrm{R}} = (2 \sqrt{2} \, T_{\pi})^{-1}$ is the Rabi frequency for $S$ = $\frac{1}{2}$,
$f_{\mathrm{R, g}}(f) = \{(f - \delta f))^2 + f_{\mathrm{R}}^2\}^{1/2}$ with $\delta f$ = $2 g \mu_B (B_0 - B_{\mathrm{res}})$ is the generalized Rabi frequency~\cite{A61},
$\beta(f) = 2 \pi f_{\mathrm{R,g}} T_{\pi}$ is the flip angle, and 
$\mathcal{S}(f) = \delta \nu_{\mathrm{res}}/2 \pi \{(f - \delta f)^2 + (\delta \nu_{\mathrm{res}}/2)^2\}$ is assumed to be a Lorentzian.
Despite the complex form, Eq.~(\ref{Tcalc}) contains only two fitting parameters (the resonance magnetic field $B_{\mathrm{res}}$ and the impurity spin linewidth $\delta \nu_{\mathrm{res}}$),
and yet reproduces the experimental data at ($B_{\mathrm{res}}$, $\delta \nu_{\mathrm{res}}$) = (52.838~mT, 520~kHz) [solid lines in Fig.~\ref{fig4}(a)].
The extracted P1 linewidth of 520~kHz is twice broader than the NV linewidth (250~kHz), 
and is much broader than the dipolar-limited linewidth estimated from the second moment (95~kHz for the total P1 + NV density of 3.75 $\times$ 10$^{17}$~cm$^{-3}$)~\cite{A61,WRHK97},
hinting the presence of additional broadening mechanisms in this sample.
On the other hand, Fig.~\ref{fig4}(b) suggests that, while the error at the best fit is 1.54~ms$^{-1}$, the 200-kHz linewidth can be obtained with the error of 1.68~ms$^{-1}$.
Such a small difference in errors can arise, for instance, from the uncertainty in $T_1$, which is also $B_0$-dependent around the SSR due to cross relaxation.~\cite{JAJ+12,MRK+15,HKS+15,WBH+16}
For a more refined estimation of the P1 linewidth, fine-tuning of $B_0$ to the exact P1 resonance as well as a detailed measurement of $T_1$ will be helpful.
Nonetheless, the method presented here will be a powerful approach to resolve a spin spectrum when applied to external spins.

In summary, by applying both DNP and pulsed ODMR techniques to a near-surface, narrow-resonance-linewidth NV ensemble,
we have shown that a photon-shot-noise-limited magnetic sensitivity of 35~nT~Hz$^{-0.5}$, highly promising for 2D magnetic imaging, is attainable.
We have also measured $T_2$ and deduced quantitative information on residual paramagnetic impurities in the sample.
Decoherence spectroscopy as demonstrated here is applicable to detect magnetic signals external to the sample,
providing a versatile tool for DC magnetic sensing.
Although the present work focused on the internal P1 spins for the purpose of demonstrating the power of decoherence spectroscopy,
the magnetic field can be readily tuned to avoid the P1 resonances, while still maintaining DNP.
Such a condition is suitable to concurrently perform highly sensitive DC magnetic imaging and decoherence spectroscopy of external spins.

W.D.L. acknowledges support from NSF of China Grant No.~61306123 and RGC of HKSAR Grant No.~27205515.
H.W. acknowledges the support from JSPS Grant-in-Aid for Scientific Research (KAKENHI) (A) No.~26249108 and JST Development of Systems and Technologies for Advanced Measurement and Analysis (SENTAN).
K.-M.C.F. acknowledges support by the National Science Foundation under Grant No.~1607869 and RCSA's Cottrell Scholar program.
K.M.I acknowledges support from KAKENHI (S) No.~26220602, JSPS Core-to-Core Program, and Spintronics Research Network of Japan (Spin-RNJ).
\bibliography{e1304}
\end{document}